\documentclass[reqno]{amsart}

\usepackage{latexsym,pifont}
\usepackage{epsfig}

\usepackage{rotating}

\usepackage{ifpdf}
\ifpdf
\usepackage{epstopdf}
\usepackage{hyperref}
\else
\usepackage[hypertex]{hyperref}
\fi

\usepackage{amsfonts,amsmath,amssymb,mathrsfs,extarrows,MnSymbol}
\usepackage[all]{xy}

\usepackage{tikz}
\usetikzlibrary{matrix,arrows}

\newcommand{\brem}{\begin{Rem}}
\newcommand{\erem}{\end{Rem}\medskip}
\newcommand{\beg}{\begin{Eg}}
\newcommand{\eeg}{\end{Eg}}

\newcommand{\barr}{\begin{array}}
\newcommand{\earr}{\end{array}}
\newcommand{\ben}{\begin{enumerate}}
\newcommand{\een}{\end{enumerate}}
\newcommand{\bit}{\begin{itemize}}
\newcommand{\eit}{\end{itemize}}

\newcommand{\qq}{\begin{eqnarray}}
\newcommand{\qqq}{\end{eqnarray}}

\newcommand{\nn}{\nonumber}

\newcommand{\ovl}[1]{\overline{#1}}

\newcommand{\Rcite}[1]{Ref.\,\cite{#1}}

\newcommand\void[1]{}

\newcommand{\tx}[1]{\textrm{#1}} 

\newcommand{\gt}[1]{\mathfrak{#1}}

\def\cA{\mathcal{A}}

\def\cG{\mathcal{G}}

\def\ceL{\mathcal{L}}

\def\cO{\mathcal{O}}

\def\xcF{\mathscr{F}}

\def\xcI{\mathscr{I}}

\def\xcK{\mathscr{K}}
\def\xcL{\mathscr{L}}

\def\xcV{\mathscr{V}}
\def\xcW{\mathscr{W}}

\def\bN{{\mathbb{N}}}

\def\bR{{\mathbb{R}}}

\def\bZ{{\mathbb{Z}}}

\def\a{\alpha}

\def\g{\gamma}
\def\G{\Gamma}

\def\D{\Delta}
\def\ep{\epsilon}
\def\vep{\varepsilon}

\def\la{\lambda}

\def\om{\omega}
\def\Om{\Omega}

\def\si{\sigma}
\def\Si{\Sigma}

\def\Bgt{\gt{B}}

\def\ggt{\gt{g}}

\def\Igt{\gt{I}}

\def\Kgt{\gt{K}}

\def\Xgt{\gt{X}}

\newcommand{\sfd}{{\mathsf d}}

\newcommand{\sfE}{{\mathsf E}}

\newcommand{\sfi}{{\mathsf i}}

\newcommand{\sfL}{{\mathsf L}}

\newcommand{\sfP}{{\mathsf P}}

\newcommand{\sfT}{{\mathsf T}}

\newcommand{\txA}{{\rm A}}

\newcommand{\txE}{{\rm E}}

\newcommand{\txg}{{\rm g}}
\newcommand{\txG}{{\rm G}}

\newcommand{\txH}{{\rm H}}

\newcommand{\txK}{{\rm K}}

\def\vH{\check{H}}

\def\id{{\rm id}}
\newcommand{\pr}{{\rm pr}}

\def\dim{{\rm dim}}

\newcommand{\Id}{{\rm Id}}

\def\bgrb{\gt{BGrb}}

\newcommand{\pLie}[1]{\,{-\hspace{-8pt}\xcL}_{#1}}
\def\p{\partial}

\newcommand{\Diff}{{\rm Diff}}

\def\Hol{{\rm Hol}}

\def\bd1{{\boldsymbol{1}}}
\def\brd0{{\boldsymbol{0}}}

\def\Ad{{\rm Ad}}

\def\Lie{{\rm Lie}}

\newcommand{\uj}{{\rm U}(1)}

\def\x{\times}
\def\ox{\otimes}

\def\lx{{\hspace{-0.04cm}\ltimes\hspace{-0.05cm}}}

\begin{document}

\title{Gauge Defect Networks in Two-Dimensional CFT${}^\dagger$}

\author{Rafa\l ~R.~Suszek${}^*$}
\address{\emph{Address:}
Katedra Metod Matematycznych Fizyki, Wydzia\l ~Fizyki Uniwersytetu
Warszawskiego, ul.\,Ho\.za 74, PL-00-682 Warszawa, Poland}
\email{suszek@fuw.edu.pl}\thanks{\hspace{-.5cm} ${}^*$ The author's
work was funded from the Polish Ministry of Science and Higher
Education grant
No.\,N N201 372736. \\
${}^\dagger$ The article is the author's contribution to the
Proceedings of the XXIX International Colloquium on
Group-Theoretical Methods in Physics (20-26 August 2012, Tianjin,
China).}

\begin{abstract}
An interpretation of the gauge anomaly of the two-dimensional
multi-phase $\si$-model is presented in terms of an obstruction to
the existence of a topological defect network implementing a local
trivialisation of the gauged $\si$-model.
\end{abstract}


\maketitle

\tableofcontents

\section{Introduction}

The Symmetry Principle and the Gauge Principle figure among the most
fundamental ideas of local QFT. In particular, they serve to
distinguish theories with a non-anomalous realisation of classical
symmetries, admitting descent to the space of orbits of an action of
the symmetry group on the space of states. This note gives an
account of recent progress in realising the Principles in 2d CFT in
the mixed geometro-cohomological framework developed by Gaw\c{e}dzki
{\it et al.} and adapted in
Refs.\,\cite{Runkel:2008gr,Suszek:2011hg} to the study of CFT
defects. As we shall argue, it is through application of the ensuing
correspondence between defects and CFT dualities along the lines of
\Rcite{Suszek:2012ddg} that the categorial construction of
Refs.\,\cite{Gawedzki:2008um,Gawedzki:2010rn,Gawedzki:2012fu} of the
gauged CFT is justified and brought to completion. Our discussion
emphasises the naturalness of Gaw\c{e}dzki's approach and paves the
way towards a rigorous description of T-duality and T-folds.
\bigskip

\section{The multi-phase $\si$-model}

The mono-phase 2d non-linear $\si$-model is a theory of maps $\,X\in
C^1(\Si,M)\,$ from a compact closed 2d metric manifold
$\,(\Si,\eta)\,$ to a metric manifold $\,(M,\txg)\,$ supporting a
bundle gerbe $\,\cG$.\ The gerbe, as defined in
Refs.\,\cite{Murray:1994db,Murray:1999ew}, is a geometric
realisation of an integral class
\qq\nn
[\tfrac{1}{2 \pi}\,\txH]\in H^3(M)\,.
\qqq
It gives rise to a Cheeger--Simons differential character
$\,\Hol_\cG(X)\,$ defined as the image of $\,[X^*\cG]\,$ under the
isomorphism $\,\vH^2\left(\Si,\uj\right) \xrightarrow{\cong}\uj$.\
The theory is determined by the Principle of Least Action applied to
the functional
\qq\nn
S_\si[X;\eta]:=-\tfrac{1}{2}\,\tint_\Si\,\txg(\sfd X
\overset{\wedge}{,}\star_\eta\sfd X)-\sfi\,\log\Hol_\cG(X)
\qqq
in which the Hodge operator $\,\star_\eta\,$ acts on the first and
the metric $\,\txg\,$ contracts the second tensor factor in $\,\sfd
X(\si)=\p_a X^\mu(\si)\,\sfd\si^a \ox\p_\mu\vert_{X(\si)}\,$ (in
local coordinates $\,\{\si^a\}^{a\in \{1,2\}}\,$ on $\,\Si\,$ and
$\,\{X^\mu\}^{\mu\in \ovl{1,\dim M}}\,$ on $\,M$). The class
$\,[\frac{1}{2\pi}\,\txH]$,\ necessary for the cancellation of the
Weyl anomaly, is represented by a 2-cocycle in the Deligne
hypercohomology of $\,M$,\ and it is in this guise that gerbes first
entered the description of the $\si$-model in
Refs.\,\cite{Alvarez:1984es,Gawedzki:1987ak}, leading to a
classification of field theories with fixed $\,(M,\txg,\txH)\,$ by
elements of $\,H^2\left(M,\uj \right)$,\ and to a geometric
quantisation of the $\si$-model through transgression in
\Rcite{Gawedzki:1987ak}.

Drawing inspiration from models of spin lattices with lines of spin
frustration and from orbifold string theory, we are led to consider,
after \Rcite{Fuchs:2007fw}, mappings from $\,\Si\,$ to a disjoint
union of manifolds $\,M$,\ with discontinuities along lines
$\,\ell\subset\Si\,$ at which the component of the energy-momentum
current generating diffeomorphisms of $\,\Si\,$ that preserve
$\,\ell\,$ is continuous, whence the name conformal defects. Amidst
these, we find topological defects with the property that all
components of the current are continuous at $\,\ell$.\ A formal
path-integral argument of \Rcite{Runkel:2008gr} shows that in a
factorisable CFT with defect lines we are bound to consider
intersections of the latter. Together with defect lines, they
compose an oriented graph $\,\G\subset\Si\,$ that splits $\,\Si\,$
into patches. Thus, we arrive at the definition of the multi-phase
$\si$-model of \Rcite{Runkel:2008gr}: It is a theory of $C^1$-smooth
maps $\,X\,$ that send the patches to a metric space $\,(M,\txg)$,\
defect lines to a smooth space $\,Q\,$ with smooth maps $\,\iota_1,
\iota_2:Q\to M\,$ and $\,\om\in\Om^2 (Q)$,\ and defect junctions of
valence $\,n\geq 3\,$ to a smooth space $\,T_n\,$ with smooth maps
$\,\pi_n^{k,k+1}: T_n\to Q,\ k\in\bZ/n\bZ\,$ subject to additional
consistency constraints. The target space
\qq\nn
\xcF:=M \sqcup Q\sqcup T\,,\qquad T:=\bigsqcup_{n\geq 3}\,T_n
\qqq
is the base of a geometric structure $\,\Bgt\,$ from the 2-category
$\,\bgrb^\nabla(\xcF)$,\ introduced and elaborated in
Refs.\,\cite{Stevenson:2000wj,Waldorf:2007mm}, of abelian bundle
gerbes with connection over $\,\xcF$.\ The structure consists of a
0-cell (gerbe) $\,\cG$,\ a 1-cell (1-isomorphism)
\qq\nn
\Phi\ :\ \iota_1^*\cG \xrightarrow{\cong}\iota_2^*\cG\ox I_\om
\qqq
(for $\,I_\om\,$ the trivial gerbe of curving $\,\om$), and 2-cells
(2-isomorphisms)
\qq\nn
\varphi_n:\circ_{k\in\bZ/n\bZ}\,\pi_n^{k,k+1\,*}\left(
\Phi^{\vep_n^{k,k+1}}\ox\Id\right)\xLongrightarrow{\cong}\Id
\qqq
(with $\vep_n^{k,k+1}=\pm 1\,$ for an incoming/outgoing defect line
with index $\,k$). The triple $\,(\cG,\Phi,\varphi_n)\,$ is a
geometric realisation of the $2\pi$-integral class of the relative
3-form
\qq\nn
\Theta:=\txH\oplus\om\oplus 0
\qqq
in the cohomology of the complex of cochain groups
\qq\nn
C^p_{\rm dR}(M,Q,T_n\vert\D_Q,\D_{T_n}):=\Om^p(M)\oplus\Om^{p-1}(Q)
\oplus\Om^{p-2}( T_n)\,,\qquad p\in\bN
\qqq
(here, $\,\Om^{-1}(X):=\bR^{\pi_0(X)}\,$ and $\,\Om^{m<-1}( X)\equiv
0$) endowed with the differential
\qq\nn
\widetilde\sfd(\om_M\oplus\om_Q\oplus\om_{T_n}):=\sfd\om_M\oplus
(-\sfd\om_Q-\D_Q\om_M)\oplus(\sfd\om_{T_n}-\D_{T_n}\om_Q)\,,
\qqq
written for $\,\D_Q:=\iota_2^*- \iota_1^*\,$ and
$\,\D_{T_n}:=\sum_{k\in\bZ /n\bZ}\,\vep_n^{k,k+1}\,\pi_n^{k,k+1
\,*}$,\ {\it cf} \Rcite{Suszek:2012ddg}. The said theory is
determined by the Principle of Least Action applied to the
functional, given in \Rcite{Runkel:2008gr},
\qq\nn
\ \ S_\si[(X\vert\G);\g]:=-\tfrac{1}{2}\,\tint_\Si\,\txg(\sfd X
\overset{\wedge}{,}\star_\g\sfd X)-\sfi\,\log\Hol_{(\cG,\Phi,
\varphi_n\,\vert\,n\in\bN_{\geq 3})}(X\vert\G)\,,
\qqq
in which the last term is a differential character expressed in
terms of Deligne data of the $\,(\cG,\Phi,\varphi_n)$.\ The
definition enables us to label conformal defects with fixed
$\,(M,\txg,\cG,Q,\iota_\a,\om)\,$ by classes in $\,H^1\left(Q,\uj
\right)$,\ and (inequivalent) choices of the $\,\varphi_n\,$ by
elements of $\,\uj^{\pi_0(T_n)}$.\ It also determines a geometric
quantisation of (the twisted sector of) the $\si$-model through a
relative variant of transgression described in
Refs.\,\cite{Gawedzki:2002se,Gawedzki:2003pm,Gawedzki:2004tu,Suszek:2011hg}.

\section{Defects vs.\ dualities}

A relation between topological defects and CFT dualities has been
predicted within the categorial approach to the quantisation of the
CFT developed by Fr\"ohlich, Fuchs, Schweigert, Runkel {\it et al.}
in Refs.\,\cite{Frohlich:2004ef,Frohlich:2006ch} and subsequent
works. Its realisation in the geometric framework outlined above was
worked out in the author's recent paper \cite{Suszek:2011hg},
preceded by a case study in \Rcite{Runkel:2008gr} of a class of
defects in the Wess--Zumino--Witten--Gaw\c{e}dzki $\si$-model of
Refs.\,\cite{Gawedzki:1987ak,Gawedzki:2002se,Gawedzki:2003pm,Gawedzki:2004tu}.
It stems from the simple world-sheet intuition: A (space-like)
circular defect $\,\ell_\circ\,$ separating two patches of the
$\si$-model establishes a correspondence between states from the
respective phases, represented by the Cauchy data localised at the
circular components of the boundary of the patches joining at
$\,\ell_\circ$.\ A canonical analysis of the relation between the
two sets of Cauchy data that follows from the gluing condition
imposed at $\,\ell_\circ$,\ as part of the variational principle,
upon classical ({\it i.e.}, minimal) field configurations $\,X\,$
demonstrates that the subspace
\qq\nn
\Igt_\si\subset\sfP_\si^{\x 2}
\qqq
composed of pairs of states in cross-defect correspondence is
isotropic with respect to $\,\pr_1^*\Om_\si-\pr_2^*\Om_\si\,$ (here,
the $\,\pr_\a:\sfP_\si^{\x 2}\to\sfP_\si\,$ are the canonical
projections). The defect data induce a full-fledged (pre)quantum CFT
duality, that is $\,\Igt_\si\,$ is a graph of a symplectomorphism
preserving the Hamiltonian density and lifting to an isomorphism
\qq\nn
\pr_1^*\ceL_\si\vert_{\Igt_\si}\cong\pr_2^* \ceL_\si\vert_{\Igt_\si}
\qqq
between pullbacks of the prequantum bundle $\,\ceL_\si\,$ of the
mono-phase $\si$-model, iff the maps $\,\sfL Q\to\sfL
M:\phi\mapsto\iota_\a\circ\phi\,$ are surjective submersions, the
defect is topological and extra technical conditions are satisfied,
{\it cf} \Rcite{Suszek:2011hg}. This is the case, in particular,
when the defect carries the data of a 1-isomorphism
$\,\Phi_F:\pr_1^*\cG\xrightarrow{\cong}\pr_2^*\cG\,$ over the graph
$\,Q=(\id_M\x F)(M)\,$ of an isometric diffeomorphism $\,F\,$ of
$\,(M,\txg)$.\ The associated duality relates mappings into a single
component of $\,M$,\ and so it represents a global symmetry of the
mono-phase $\si$-model. This brings us to the main point of our
discussion.

\section{The Gauge Principle through defect networks}

Global symmetries of the $\si$-model form a subgroup
\qq\nn
\txG_\si \subset\Diff(\xcF)
\qqq
acting on $\,\xcF\,$ as
\qq\nn
\ell\ :\ \txG_\si\x \xcF\to\xcF\,,
\qqq
with the $\,\iota_\a\,$ and $\,\pi_n^{k,k+1}\,$ assumed
$\txG_\si$-equivariant. Elements of $\,\txG_\si\,$ satisfy the
defining relation
\qq\nn
S_\si[(\ell_g(X)\vert\G );\eta]=S_\si[(X\vert \G);\eta ]\,.
\qqq
In order to understand the underlying geometry, it is instructive to
examine infinitesimal symmetry transformations, induced by flows
$\,\psi_\cdot:]-\ep,\ep[ \x\xcF\to\xcF,\ \ep>0\,$ of the
$\txG_\si$-fundamental vector fields $\,\xcK\in\Xgt(\xcF)$.\
Inspection of the identity
\qq\nn
\tfrac{\sfd \ }{\sfd t}\big\vert_{t=0}S_\si[(\psi_t\circ
X\vert\G);\eta]=0
\qqq
reveals that the symmetries correspond to those sections
\qq\nn
\xcK\oplus \txK\in\Xgt(\xcF)\oplus C^1_{\rm
dR}(M,Q,T_n\vert\D_Q,\D_{T_n} )
\qqq
of the generalised tangent bundles (here, $\,\G\left((
T_n\x\bR)_0\right)=\bR^{\pi_0(T_n)}$)
\qq\nn
\txE\xcF:=(\sfT M\oplus\sfT^*M)\sqcup\left(\sfT Q\oplus(Q\x\bR)
\right)\sqcup\bigsqcup_{n\geq 3}\,\left(\sfT T_n\oplus(T_n\x\bR)_0
\right)\to\xcF
\qqq
which obey the relations
\qq\nn
\pLie{\xcK}\txg=0\qquad\tx{and}\qquad\widetilde
\iota_\xcK\Theta=-\widetilde\sfd\txK\,,
\qqq
expressed in terms of the relative contraction
\qq\nn
\widetilde\iota_\xcV(
\om_M\oplus\om_Q\oplus\om_{T_n}):=\iota_\xcV\om_M\oplus(-\iota_\xcV
\om_Q)\oplus\iota_\xcV\om_{T_n}\,,
\qqq
{\it cf} \Rcite{Suszek:2012ddg}. There is an essentially unique
extension of the Lie bracket on $\,\Xgt(\xcF)\,$ to $\,\G(\txE\xcF
)\,$ that closes on the set $\,\G_\si(\txE\xcF)\,$ of these
sections, to wit,
\qq\nn
[\xcV\oplus\upsilon,\xcW\oplus\varpi]_{\rm
C}^\Theta:=[\xcV,\xcW]\oplus
[\pLie{\xcV}\varpi-\pLie{\xcW}\upsilon-\tfrac{1}{2}\,\widetilde
\sfd(\widetilde\iota_\xcV\varpi-\widetilde\iota_\xcW\upsilon)+
\widetilde\iota_\xcV\widetilde\iota_\xcW\Theta]\,,
\qqq
and it is natural to think of it as a relative $\Theta$-twisted
Courant bracket. Indeed, it restricts to the $\txH$-twisted Courant
bracket on sections of the standard generalised tangent bundle
$\,\sfT M\oplus\sfT^*M\,$ whose r\^ole in the description of
symmetries of the mono-phase $\si$-model has been known since
\Rcite{Alekseev:2004np}. The intrinsically gerbe-theoretic nature of
the bracket is readily established with the help of Hitchin-type
isomorphisms of \Rcite{Suszek:2012ddg}. With the algebraic structure
on $\,\G_\si(\txE\xcF)\,$ thus determined, we are ready to discuss
the Gauge Principle.

Geometrically, the Principle stipulates that the covariant
configuration bundle $\,\xcF':=\Si\x\xcF\,$ of the $\si$-model be
replaced by the bundle $\,\sfP\x_\ell\xcF\,$ associated, through
$\,\ell$,\ to an arbitrary principal $\txG_\si$-bundle
$\,\sfP\to\Si\,$ and carrying a fibrewise action of the adjoint
bundle $\,\sfP\x_\Ad \txG_\si$.\ In this manner, the action of the
symmetry group (encoded in the isotype $\,\txG_\si\,$ of the fibre
of $\,\sfP \x_\Ad\txG_\si$) is rendered local while preserving the
original field content, the latter being determined by the isotype
$\,\xcF\,$ of the fibre of $\,\sfP\x_\ell\xcF$.\ Clearly, we must
insist that the new covariant configuration bundle admit a
\emph{global} section, to be identified with the lagrangean field of
the gauged $\si$-model under construction. The geometric descent
from the extended covariant configuration bundle
$\,\widetilde\xcF:=\sfP\x\xcF\,$ to $\,\sfP\x_\ell\xcF\,$ is
straightforward as the $\txG_\si$-action on the former is free. The
nontrivial task is to lift this action equivariantly to an extension
$\,\widetilde\Bgt_\cA\,$ of the original $\,\Bgt\,$ to
$\,\widetilde\xcF\,$ obtained by coupling its pullback to the gauge
field given by the principal $\txG_\si$-connection 1-form
\qq\nn
\cA\in\Om^1(\sfP)\ox\ggt_\si\,,\qquad\ggt_\si:=\Lie\,\txG_\si
\qqq
on $\,\sfP$.\ This is to be done in such a manner that $\,\widetilde
\Bgt_\cA\,$ descends to the quotient $\,\sfP\x_\ell\xcF\,$ in the
sense captured by a variant, established in
Refs.\,\cite{Gawedzki:2008um,Gawedzki:2010rn,Gawedzki:2012fu}, of
Stevenson's Principle of Descent of
Refs.\,\cite{Stevenson:2000wj,Waldorf:2007mm} (to be viewed as a
(2-)categorification of pullback) and defines a CFT invariant under
the action of the gauge group $\,\sfP\x_\Ad \txG_\si$.

It is convenient to begin the analysis with the topologically
trivial case $\,\sfP=\Si\x\txG_\si$,\ in which we keep $\,\xcF'\,$
as the covariant configuration bundle and replace $\,\cA\,$ by the
connection 1-form $\,\txA=\txA^A\ox t_A\in\Om^1(\Si)\ox \ggt_\si$,\
written in terms of the generators $\,t_A\,$ of $\,\ggt_\si\,$
subject to the relations
\qq\nn
[t_A,t_B]= f_{ABC}\,t_C
\qqq
in which the $\,f_{ABC}\,$ are the structure constants of
$\,\ggt_\si$.\ Upon choosing a basis
\qq\nn
\Kgt_A:=\xcK_A\oplus( \kappa_A\oplus k_A\oplus 0)\,,\qquad
A\in\ovl{1, \dim\,\ggt_\si}
\qqq
in $\,\G_\si(\sfE\xcF)\,$ projecting to the set of
$\txG_\si$-fundamental vector fields $\,\xcK_A\,$ associated with
the $\,t_A$,\ we obtain the following (minimal) extension
$\,\Bgt_\txA'\,$ of $\,\Bgt\,$ to $\,\xcF'\,$ (subscripts 1,2 denote
pullbacks along the canonical projections) given in
Refs.\,\cite{Gawedzki:2008um,Gawedzki:2010rn,Gawedzki:2012fu}: the
metric
\qq\nn
\txg_\txA:=\txg_2-\txg(\xcK_A,\cdot)_2\ox\txA^A_1-\txA^A_1\ox\txg(
\xcK_A,\cdot)_2+\txg(\xcK_A,\xcK_B)_2\,\txA^A_1\ox\txA^B_1
\qqq
and the 0-cell
\qq\nn
\cG_\txA :=\cG_2\ox I_{\rho_\txA}\qquad\tx{with}\qquad
\rho_\txA:=\kappa_{A\,2}\wedge\txA^A_1-
\frac{1}{2}\,(\iota_{\xcK_A}\kappa_B)_2\,\txA^A_1\wedge\txA^B_2
\qqq
over $\,M'$;\ the 1-cell
\qq\nn
\Phi_\txA:=\Phi_2\ox J_{\la_\txA}
\qqq
with the trivial 1-cell (trivial line bundle) $\,J_{\la_\txA}\,$
with a connection 1-form
\qq\nn
\la_\txA:=-k_{A\,2}\,\txA^A_1
\qqq
over $\,Q'$;\ the 2-cells
\qq\nn
\varphi_{n\,\txA}:=\varphi_{n\,2}
\qqq
over the $\,T_n'$.\ In terms of these, we define -- for $\,\xi:=(
\id_\Si,X)\,$ with $\,X\,$ as before -- the gauged $\si$-model
coupled to the topologically trivial gauge field
\qq\nn
S_\si[(X\vert\G);\txA,\g]:=-\tfrac{1}{2}\,\tint_\Si\,\txg_\txA(\sfd
\xi\overset{\wedge}{,}\star_\g\sfd\xi)-\sfi\,\log\Hol_{(\cG_\txA,
\Phi_\txA,\varphi_{n\,\txA}\,\vert\,n\in\bN_{\geq 3})}(\xi\vert\G)\,
\qqq
We may next impose the demand that the above be invariant under
gauge transformations
\qq\nn
(X,\txA)\mapsto(\ell_\chi(X),\Ad_\chi \txA-\sfd\chi\,\chi^{-1})
\qqq
defined for arbitrary $\,\chi\in C^\infty(\Si,\txG_\si)$.\ For
$\,\chi\,$ homotopic to the identity, this yields the condition that
the triple
\qq\nn
(\bigoplus_{A=1}^{\dim\,\ggt_\si}\,C^\infty(\xcF,\bR)\,\Kgt_A,[\cdot,
\cdot]_{\rm C}^\Theta,\a_{\sfT\xcF}\equiv\pr_1)
\qqq
define a Lie algebroid, and the latter is then canonically
isomorphic with the action algebroid $\,\ggt_\si\lx_\ell\xcF$,\ {\it
cf} Refs.\,\cite{Gawedzki:2008um,Gawedzki:2010rn,Gawedzki:2012fu}.
Invariance under large gauge transformations requires the existence
of a 1-cell
\qq\nn
\Upsilon\ :\ \ell^*\cG \xrightarrow{\cong}\pr_2^*\cG\ox
I_{\rho_{\theta_L}}
\qqq
(for $\,\theta_L\,$ the left-invariant Maurer--Cartan 1-form on
$\,\txG_\si$) and a 2-cell
\qq\nn
\Xi\ :\ \ell^*\Phi\xLongrightarrow{\cong}
[((\id_{\txG_\si}\x\iota_2)^*\Upsilon^{-1}\ox\Id)
\circ(\pr_2^*\Phi\ox\Id)\circ(\id_{\txG_\si}\x\iota_1)^*\Upsilon
]\ox J_{\la_{\theta_L}}
\qqq
(subject to additional constraints) from the 2-category
$\,\bgrb^\nabla(\txG_\si\x\xcF)\,$ based on the arrow manifold of
the action groupoid $\,\txG_\si\lx_\ell\xcF$.\ Upon recalling that
$\,\ggt_\si\lx_\ell\xcF\,$ is the tangent algebroid of
$\,\txG_\si\lx_\ell\xcF$,\ we are led to identify the latter as the
algebro-differential structure underlying those symmetries of the
$\si$-model that can be gauged. This observation was elucidated in
\cite{Suszek:2012ddg} with the help of the equivalence
\qq\nn
\txG_\si\textrm{-}\gt{Bun}(\Si\Vert\xcF)\cong\txG_\si\lx_\ell
\xcF\textrm{-}\gt{Bun}( \Si)
\qqq
between the groupoid of principal $\txG_\si$-bundles $\,\sfP
\to\Si\,$ with the property that $\,\sfP\x_\ell\xcF\,$ admits a
global section and the groupoid of principal $\txG_\si\lx_\ell
\xcF$-bundles over $\,\Si$,\ the latter being defined in
\Rcite{Moerdijk:1991}. The equivalence has a natural world-sheet
interpretation, reminiscent of the categorial generalised-orbifold
construction of \Rcite{Frohlich:2009gb}: An object $\,\sfP\,$ of
$\,\txG_\si\textrm{-} \gt{Bun}(\Si\Vert\xcF)\,$ (data of the gauged
$\si$-model, forgetting the connection) trivialising over an open
cover $\,\cO:=\{\Si_i\}_{i\in\xcI}\,$ of $\,\Si\,$ corresponds to a
family of smooth mappings $\,X_i:\Si_i\to\xcF\,$ under which images
of points $\,\si\in\Si_i\cap\Si_j\,$ are related as
$\,X_i(\si)=\ell_{g_{ij}}(X_j)(\si)\,$ by arrows of $\,\txG_\si
\lx_\ell\xcF\,$ defined in terms of transition maps $\,g_{ij}\,$ of
$\,\sfP$.\ This demonstrates the necessity of incorporating
nontrivial principal $\txG_\si$-bundles in a complete description of
the gauged $\si$-model as they are necessary to reproduce the
twisted sector of the descended CFT. Their presence appears to
ensure self-consistency of the quantised theory, {\it cf}
Refs.\,\cite{Gawedzki:2008um,Gawedzki:2010rn,Gawedzki:2012fu}.

The goal of extending the Gauge Principle to topologically
nontrivial bundles $\,\sfP\,$ can be attained through local
trivialisation of $\,\sfP\,$ over $\,\cO\,$ followed by the gluing
of local data $\,(X_i,\txA_i)\,$ over the $\,\Si_i\,$ and of the
local phases $\,\Bgt_{\txA_i}'\,$ of the gauged $\si$-model by means
of the $\,g_{ij}\,$ at the edges $\,\ell_{ij}\subset\Si_i
\cap\Si_j\,$ of a trivalent graph $\,\G_\cO\,$ associated with
$\,\cO$,\ with $\,\G_\cO\cap\G\,$ discrete and not containing
vertices of $\,\G$.\ We then check, as in \Rcite{Suszek:2012ddg},
that the pair $\,(\Upsilon,\Xi)$,\ required to exist by previous
arguments, can be used to induce the desired structure of a
topological defect at the $\,\ell_{ij}$,\ alongside that of a
(trans-)defect junction at $\,\G_\cO\cap\G$.\ These jointly
implement the gluing of local data in the presence of $\,\G$,\ and
the gluing itself emerges as a duality between the trivialised
phases $\,\Bgt_{\txA_i}'$.\ The hitherto assignments already fix, by
the arguments from \Rcite{Runkel:2008gr}, the structure to be pulled
back to the vertices of $\,\G_\cO$,\ and so we are led to demand the
existence of a 2-cell
\qq\nn
\g\ :\ (d^{(0)\,*}_2\Upsilon\ox\Id)\circ
d^{(2)\,*}_2\Upsilon\xLongrightarrow{\cong}d^{(1)\,*}_2\Upsilon
\qqq
of $\,\bgrb^\nabla(\txG_\si^2\x M)$,\ written in terms of the face
maps $\,d^{(k)}_l:\txG_\si^l\x M\to\txG_\si^{l- 1}\x M,\
k\in\ovl{0,l}\,$ of the nerve of $\,\txG_\si\lx_\ell\xcF\,$ and
playing a r\^ole analogous to that of $\,\varphi_3$.\ At this stage,
we still have to demand that the gauged $\si$-model thus sewn from
the local phases $\,\Bgt_{\txA_i}'\,$ be independent of the
arbitrary choices made, {\it e.g.}, the choice of $\,\cO\,$ and
$\,\G_\cO$,\ of the trivialisation and of index assignments. This
further constrains $\,(\Upsilon,\g,\Xi)$,\ so that ultimately we
have to demand that the triple define a $\txG_\si$-equivariant
structure on $\,\Bgt$,\ as was demonstrated in
\Rcite{Suszek:2012ddg}. By the Principle of Descent, the gauged
$\si$-model then yields a (manifestly gauge invariant) CFT on
$\,\sfP\x_\ell\xcF$,\ equivalent to a $\si$-model on
$\,\xcF/\txG_\si\,$ whenever the latter exists as a manifold, {\it
cf} Refs.\,\cite{Gawedzki:2008um,Gawedzki:2010rn,Gawedzki:2012fu}.
Obstructions to the existence of the $\txG_\si$-equivariant
structure (known as gauge anomalies), as well as its inequivalent
realisations are classified in
Refs.\,\cite{Gawedzki:2008um,Gawedzki:2010rn,Gawedzki:2012fu} by a
$\txG_\si$-equivariant extension of the relative Deligne
hypercohomology.\bigskip

\noindent{\bf Acknowledgements:} The author is grateful to
K.~Gaw{\c{e}}dzki and I.~Runkel for valuable discussions. He is also
pleased to acknowledge the kind hospitality of the Chern Institute
of Mathematics in Tianjin at the time of the XXIX International
Colloquium on Group-Theoretical Methods in Physics.

\bibliographystyle{amsalpha}

\end{document}